%% file: main.tex
\documentclass[10pt, conference]{IEEEtran}
\usepackage{graphicx, xcolor, colortbl}
\usepackage{algorithm, algpseudocode, subfigure}
\usepackage{url,amsmath,amssymb,array,threeparttable,multirow,booktabs, tabularx}
\usepackage[sort&compress,square,comma,numbers]{natbib}
\def\BibTeX{{\rm B\kern-.05em{\sc i\kern-.025em b}\kern-.08em
    T\kern-.1667em\lower.7ex\hbox{E}\kern-.125emX}}

\newcommand{\pp}{\text{$D^3$}}
\newcommand{\cc}{\text{0.984}}
\newcommand{\dt}{\text{0.961}}
\begin{document}
\title{Deep Learning-based Real-time Smartphone Pose Detection for Ultra-wideband Tagless Gate}

\author{
\IEEEauthorblockN{
Junyoung Choi, Sagnik Bhattacharya\
    }
\IEEEauthorblockA{
\textit{Communications Standards Research Team, Samsung Electronics, Seoul, South Korea} \\
Email: \{juny.choi, sagnik.b\}@samsung.com
    }
}

\maketitle

\begin{abstract}
As commercial interest in proximity services increased, the development of various wireless localization techniques was promoted.
In line with this trend, Ultra-wideband (UWB) is emerging as a promising solution that can realize proximity services thanks to centimeter-level localization accuracy.
In addition, since the actual location of the mobile device (MD) on the human body, called \textit{pose}, affects the localization accuracy, poses are also important to provide accurate proximity services, especially for the UWB tagless gate (UTG).
In this paper, a real-time pose detector, termed {\pp}, is proposed to estimate the pose of MD when users pass through UTG.
{\pp} is based on line-of-sight~(LOS) and non-LOS~(NLOS) classification using UWB channel impulse response and utilizes the inertial measurement unit embedded in smartphone to estimate the pose. 
{\pp} is implemented on Samsung Galaxy Note20 Ultra (i.e., SM-N986B) and Qorvo UWB board to show the feasibility and applicability.
{\pp} achieved an LOS/NLOS classification accuracy of {\cc}, and ultimately detected four different poses of MD with an accuracy of {\dt} in real-time.

\end{abstract}


\input{1_introduction}
\input{2_related}
\input{3_processing}
\input{4_proposed}
\input{5_evaluation}
\input{6_conclusion}

% \footnotesize
\scriptsize
\bibliographystyle{IEEEtran}
\bibliography{main.bib}

\end{document}

%% file: 1_introduction.tex
\section{Introduction}
\label{sec:intro}

Recently, Ultra-wideband (UWB) is emerging as a promising solution for proximity services with centimeter-level localization performance compared to other narrowband radios such as Wi-Fi and Bluetooth~\cite{choi2020smartphone, choi2022wand}.
UWB-based proximity services, such as UWB tagless gate (UTG)~\cite{tagless}, UWB payment~\cite{projectnear}, and UWB car key~\cite{ccc, digitalcar} are preparing to come out to the world with a lot of attention due to commercial and social values.
In addition, the flagship smartphones of Samsung (e.g., Galaxy Note 20 Ultra and later) are equipped with UWB module.

\begin{figure}[t]
    \centering
    \subfigure[Normal operation]{
    \includegraphics[width=0.29\columnwidth]{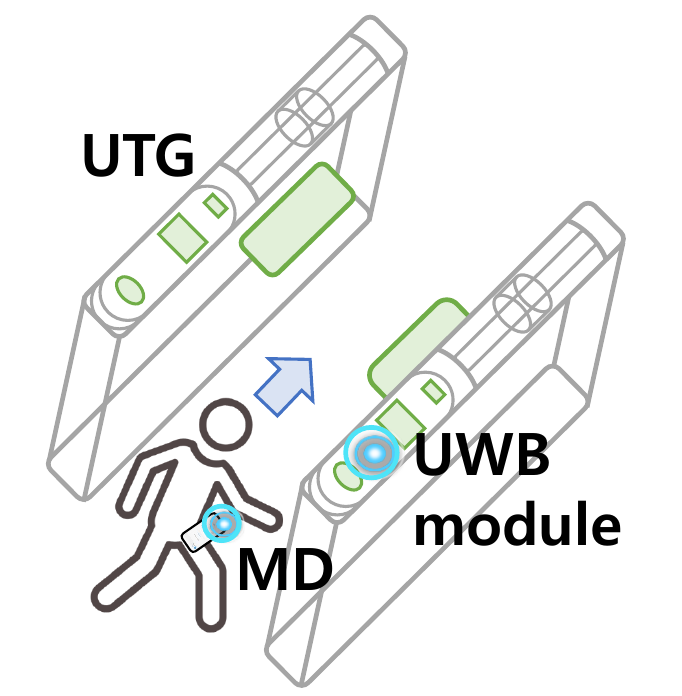}
    \label{fig:normal_operation}
    }
    \subfigure[Unexpected operations]{
    \includegraphics[width=0.58\columnwidth]{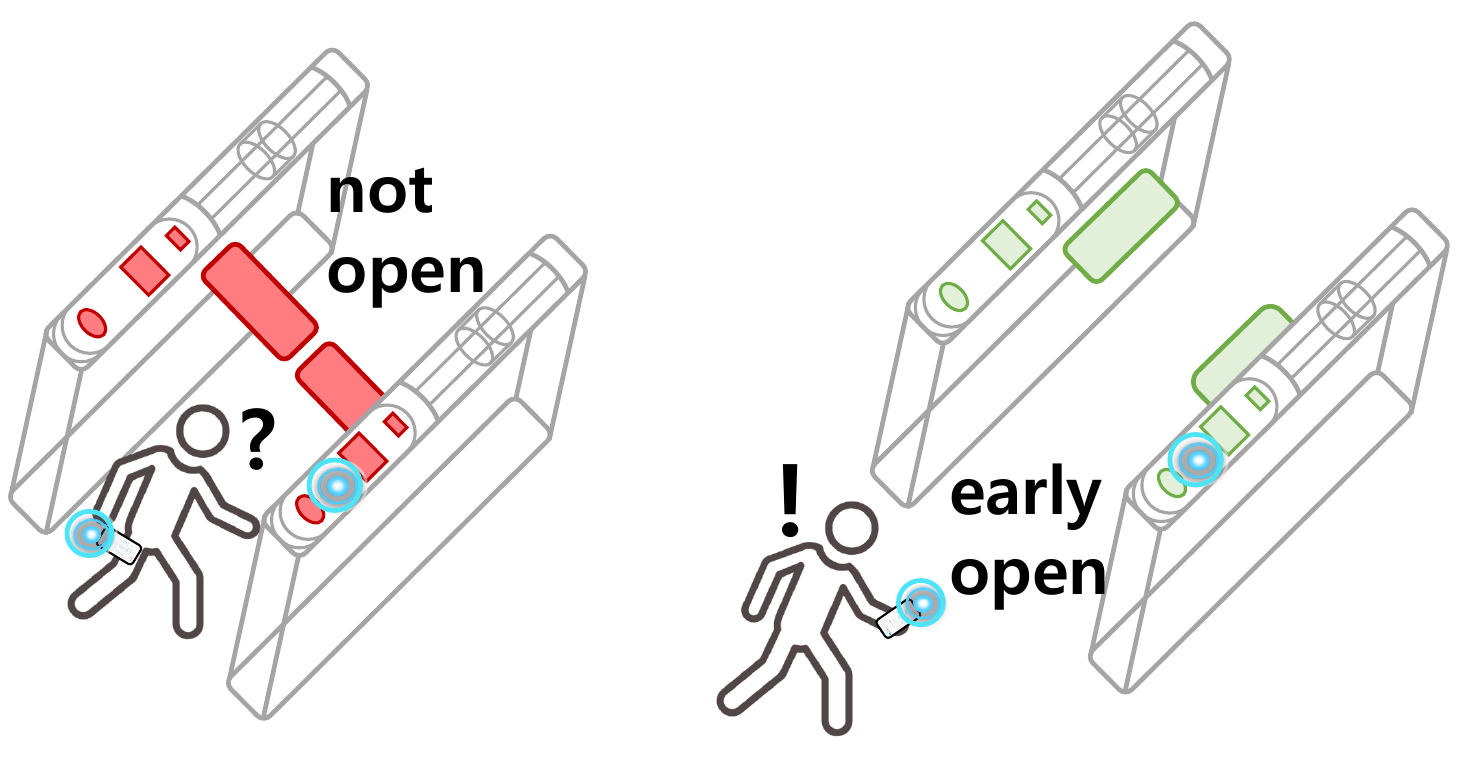}
    \label{fig:unexpected_operation}
    }
    \vspace{-2mm}
    \caption{The normal and unexpected operations of UTG. The opening and closing of the gate are colored in green and red, respectively.}
    \label{fig:scenario}
    \vspace{-5mm}
\end{figure}

The mentioned applications are based on the distance between the two UWB modules to ensure proximity, and the most common way to calculate the distance is double-sided two-way ranging (DS-TWR).
However, even if the DS-TWR-based ranging result provides a centimeter level of localization accuracy, the actual location of the user may be different from that of the mobile device (MD)~\cite{grosswindhager2018salma}.
For example, when the user approaches the UTG, the MD and the UTG perform DS-TWR to smoothly open the gate when the user enters a specific area of the gate.
In this situation, due to \textit{the position of MD on the human body}, which is called \textit{pose} in this paper, the user can experience unexpected situations
as shown in Fig.~\ref{fig:scenario}.

When the user is close enough to the gate, the gate must be automatically opened without causing any interference to the user, as shown in Fig.~\ref{fig:normal_operation}.
However, if the MD is in the back pocket, the gate might not be opened because of the incorrect ranging results due to a blockage, signal attenuation, and multipath, or the gate might be opened too early when the MD is in front of the body, as shown in Fig.~\ref{fig:unexpected_operation}.
Due to the difference between the actual location of the human body and the pose of the MD, both unexpected behaviors occur.
These unexpected operations may be resolved when the opening of the gate is adaptively operated based on the pose of the MD.

In this paper, we present {\pp}, \underline{\textbf{D}}eep learning-based pose \underline{\textbf{D}}etection of the MD on UWB \underline{\textbf{D}}S-TWR.
We newly design and implement a novel pose detection system based on UWB channel impulse response (CIR) and inertial measurement unit (IMU) sensor.
We define four different poses of the MD, and classify them with two steps using deep learning method.
% , specifically using convolutional neural network~(CNN) and long-short-term memory (LSTM).
First, {\pp} collects CIR information during DS-TWR operation and determines whether the signal is line-of-sight (LOS) or non-LOS (NLOS).
When performing LOS/NLOS classification, {\pp} sorts out effective CIR (eCIR) to minimize the input size of the CNN model, and hence, achieves real-time operation.
Finally, {\pp} determines the pose of the MD based on the LOS/NLOS classification result.

Our major contributions are summarized as follows:
\begin{itemize}
\item 
As far as we know, this is the first work to propose real-time LOS/NLOS classification and pose detection operating on Android OS with help of the UWB chipset.
We believe that our research can be an important technology to encourage UTG proximity service using UWB.

\item 
Based on deep learning models, {\pp} is designed by combining the LOS/NLOS classifier and pose detector.
To realize a real-time operation, only eCIRs are used to classify the LOS/NLOS condition to minimize the data transfer latency between the smartphone and UWB board.

\item 
We implemented {\pp} in the commercial Android OS and Qorvo UWB module, and evaluated its performance to show the feasibility.
{\pp} classified LOS/NLOS condition of four poses with an accuracy of {\cc} and achieved an accuracy of {\dt} for pose detection in real-time.
\end{itemize}

%% file: 2_related.tex
\begin{figure}[t]
\centering
\includegraphics[width=0.8\columnwidth]{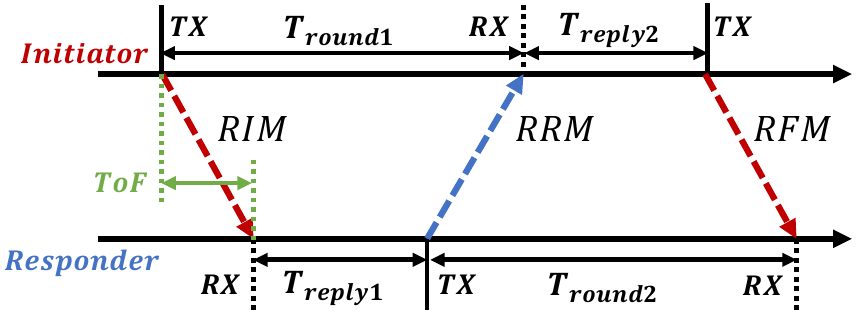}
\vspace{-2mm}
\caption{The procedure of DS-TWR with ranging messages.}
\label{fig:dstwr_procedure}
\vspace{-4mm}
\end{figure}
\section{Background and related work}
\label{sec:background_related}
\subsection{Background}
UWB is a wireless communication technique that occupies a wide bandwidth over 500~MHz, and hence, this technique uses radio frequency pulses with a very short time-duration, resulting in high time resolution.
To measure the distance between two UWB devices, the IEEE 802.15.4z standard, published in 2020, defines DS-TWR for ranging~\cite{ieee80215}. 
DS-TWR is an extension of single-sided two-way ranging in which two round-trip time measurements are used and combined to give the time-of-flight (ToF) result.
The operation of DS-TWR is shown in Fig.~\ref{fig:dstwr_procedure}.
The initiator measures the first round-trip time by sending ranging initiation message~(RIM) and the responder replies by sending ranging reply message~(RRM).
After sending RRM, the responder measures the second-trip time to which the initiator replies by sending ranging final message~(RFM).
% to complete DS-TWR exchange.
Each device precisely measures the transmission and reception times of the messages, and the resultant of ToF can be estimated by the following equation:
\begin{equation}
    {ToF} = \frac{T_{round1}T_{round2} - T_{reply1}T_{reply2}} {T_{round1} + T_{round2} + T_{reply1} + T_{reply2}}.
\end{equation}

\subsection{Related work}
To the best of our knowledge, this is the first work to implement real-time pose detection based on UWB on commercial smartphones, including LOS/NLOS classification.
The previous work focuses on LOS/NLOS classification itself or UWB error mitigation based on the classification results, and none of the previous researches have considered how to use the classification results for practical UWB application.
We are focusing on applications such as pose detection by combining IMU sensor data for proximity services as well as real-time LOS/NLOS classification.

Jiang~\emph{et al.}~\cite{jiang2020los} propose CNN-LSTM deep learning method for UWB LOS/NLOS classification based on CIR, which does not consider the input size of CIR samples for real-time operation on MD.
Jiang~\emph{et al.}~\cite{jiang2020denoise} focus on de-nosing the CIR data to identify NLOS signal based on CNN model, and test the effect of the de-nosing CIR method.
Wymeersch~\emph{et al.}~\cite{wymeersch2012machine} present the approach to directly mitigate ranging errors in both LOS and NLOS conditions.
Zeng~\emph{et al.}~\cite{zeng2018nlos} take consideration of the necessary size of CIRs among whole CIR samples, but does not conduct any implementation and real-world experiment.
Barral~\emph{et al.}~\cite{barral2019nlos} employ machine learning techniques to analyze several sets of real UWB measurements to identify the NLOS propagation condition, and conduct experiments with diverse machine learning algorithms.

%% file: 3_processing.tex
\begin{figure}[t]
\centering
\includegraphics[width=0.9\columnwidth]{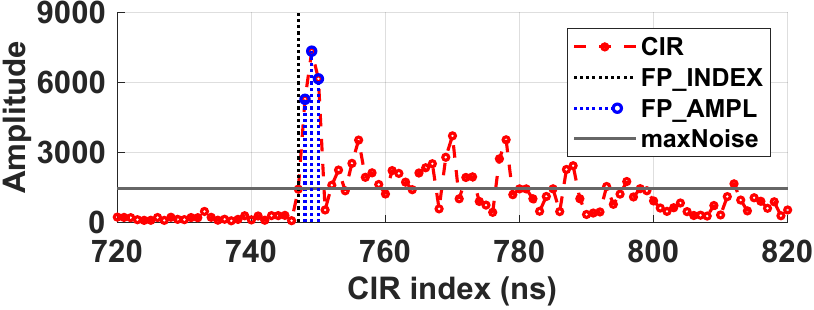}
\vspace{-2mm}
\caption{The example of CIR data including channel diagnostics. The values of FP\_INDEX and maxNoise are 747 and 1409, respectively.}
\label{fig:cir_example}
\vspace{-2mm}
\end{figure}

\begin{figure}[t]
    \centering
    \subfigure[LOS-HAND]{
    \includegraphics[width=0.21\columnwidth]{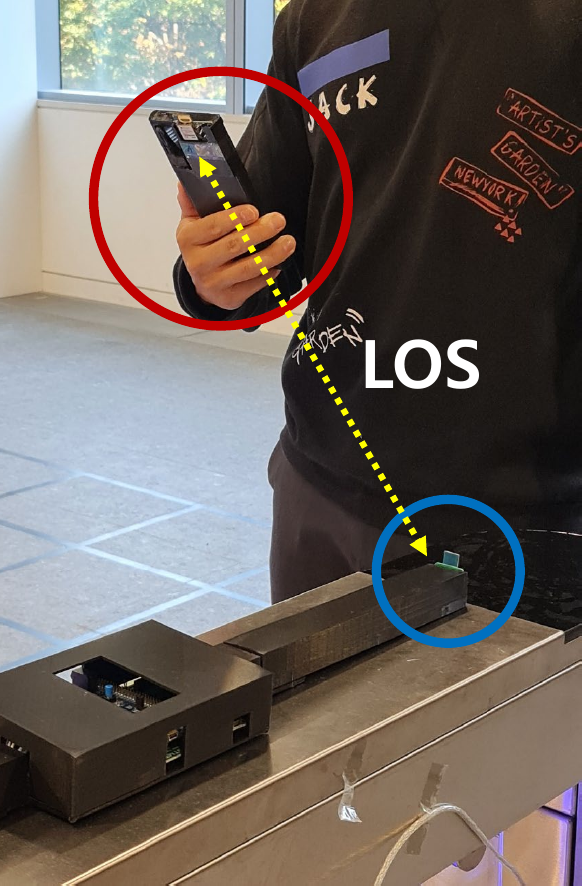}
    \label{fig:handLOS}
    }
    \subfigure[NLOS-HAND]{
    \includegraphics[width=0.21\columnwidth]{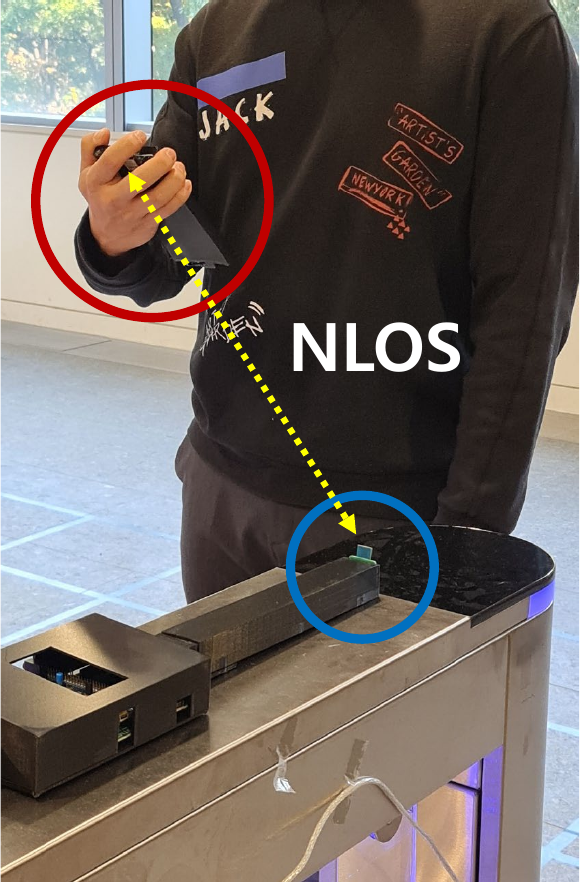}
    \label{fig:handNLOS}
    }
    \subfigure[FRONT]{
    \includegraphics[width=0.21\columnwidth]{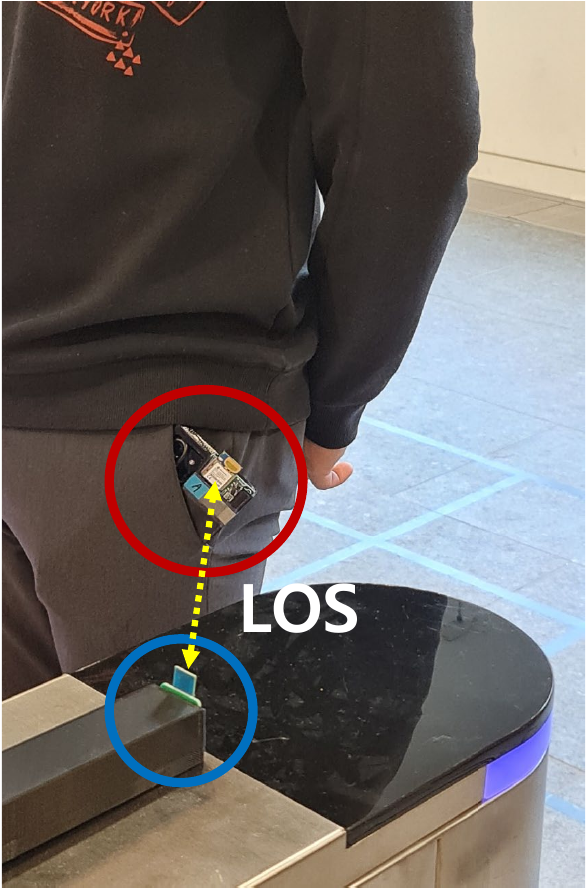}
    \label{fig:frontpocket}
    }
    \subfigure[BACK]{
    \includegraphics[width=0.21\columnwidth]{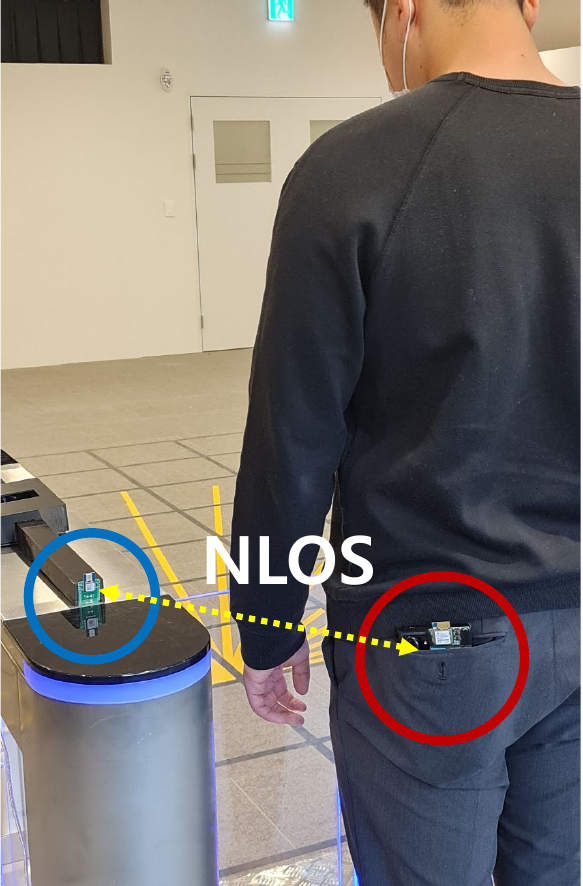}
    \label{fig:backpocket}
    }
    \vspace{-2mm}
    \caption{The illustration of four different poses for LOS and NLOS.}
    \label{fig:fourpose}
    \vspace{-4.5mm}
\end{figure}

\section{Preliminary: UWB data collection}
\label{sec:classifier}
\subsection{Channel impulse response and received signal quality}
\label{subsec:cir_diagnos}
% The UWB module in MD collects CIR during DS-TWR with UTG by receiving RFM.
We use Qorvo MDEK1001 equipped with DWM1001 UWB module to collect CIR data and the received signal quality~\cite{mdek1001}.
Both data are collected from the responder after receiving RFM during DS-TWR operation. 
First, the way to obtaining CIRs is access to the register file and calculating the complex value (i.e., a 16-bit real integer and a 16-bit imaginary integer), and then 1016 CIRs per signal can be generated from UWB module\footnote{The span of the CIR data is one symbol time. This is 992 (1016) samples for the nominal 16 (64) MHz mean pulse repetition frequency.}.
In addition, 8 kinds of information about the received signal quality include first path index (FP\_INDEX), first path amplitude (FP\_AMPL), and the maximum noise (maxNoise)\footnote{The APIs, such as dwt\_readaccdata() and dwt\_readdiagnostics(), are used for CIRs and diagnostics collection. Each CIR index interval is 1~ns}.
% using the API
% named \textbf{dwt\_readaccdata()}
% In addition, \textbf{dwt\_readdiagnostics()} provides 8 kinds of information about the received signal quality including first path index (FP\_INDEX), first path amplitude (FP\_AMPL), the maximum noise (maxNoise), etc.

Fig.~\ref{fig:cir_example} shows the value of CIR from 720 to 820 among total 1016 CIRs and three diagnostics.
% Each CIR interval is 1~ns.
The horizontal line, black stem, and blue stems represent FP\_INDEX, FP\_AMPL, and maxNoise, respectively.
The FP\_INDEX index gives the actual signal reception timing to help find the start of a meaningful CIR.
% , which helps to find the start of meaningful CIRs.
The three consecutive indices after FP\_INDEX are three FP\_AMPLs, including the signature of the received signal whether the signal is LOS or NLOS.
Based on the above information, the signals can be classified as LOS or NLOS.

\subsection{Definition of four poses of the mobile device}
We adopt deep learning model for LOS/NLOS classification, and hence, the CIR samples with the ground truth value of LOS/NLOS are essential.
In order to generate the LOS/NLOS signals, four representative poses of the MD are defined as shown in Fig.~\ref{fig:fourpose}.
Note that the pose refers to the position of the MD on the human body.
We first divide the pose into two conditions, such as LOS or NLOS, and both conditions can exist when the MD is in the hand or pocket.
Thus, each combination of $\{$LOS,~NLOS$\}\times\{$HAND,~POCKET$\}$ will be one of the poses of the MD.
First, LOS-HAND is the pose in which the user holds the MD without blocking the UWB antenna (Fig.~\ref{fig:handLOS}). 
However, in the case of NLOS-HAND, the LOS path is blocked because the user covers the UWB antenna by hand (Fig.~\ref{fig:handNLOS}).
% the user covers the UWB antenna with the hand, and hence, LOS path is blocked (Fig.~\ref{fig:handNLOS}).
If the MD is in the front pocket or back pocket, it is considered as an LOS or NLOS environment, respectively.
% we consider it as LOS or NLOS environment, respectively.
Since we are considering the scenario that the user approaches the gate, the MD can receive the LOS path directly from the gate if it is in the front pocket, which is called FRONT\footnote{We intentionally took a picture of the UWB antenna outside the front pocket to show the position of MD more clearly. During the actual experiments, the MD is completely in the front pocket.} (Fig.~\ref{fig:frontpocket}).
Conversely, when the MD is in the back pocket, the signal becomes NLOS because the human body blocks the signal, which is called BACK (Fig.~\ref{fig:backpocket}).
% will be NLOS since the human body blocks the signal, which is termed BACK (Fig.~\ref{fig:backpocket}).

%
\begin{figure}[t]
\centering
\includegraphics[width=0.9\columnwidth]{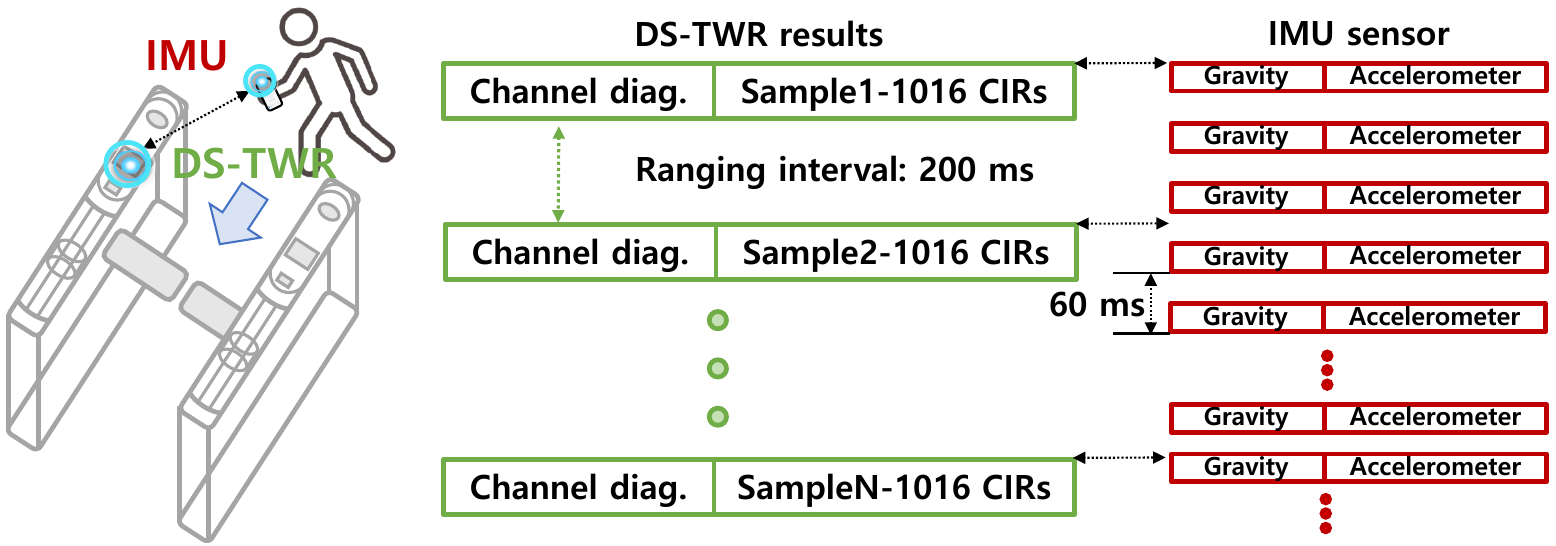}
\caption{The diagnostics/CIR and IMU data collection during DS-TWR.}
\vspace{-3mm}
\label{fig:data_collection}
\vspace{-3.5mm}
\end{figure}

\subsection{CIR and IMU data collection}
In order to train the classification model, CIR and IMU data with labels including the ground truth of LOS/NLOS and pose are required.
% The CIR and IMU data with labels, which contain the ground truth of LOS/NLOS and pose, are required to train the classification model.
Based on the four different poses of the MD presented in Fig.~\ref{fig:fourpose}, CIR and IMU data samples through preliminary experiments.
Fig.~\ref{fig:data_collection} shows the concept of data collection.
We assume that the user gets close to the gate while the MD and the gate perform DS-TWR.
The ranging interval is set to as the default value of 200~ms, 
Therefore, the MD collects the ranging results (i.e.,  1016 CIRs and a channel diagnostics) once every 200~ms.
We collect 2,000 DS-TWR results for each pose, which means that 4,000 LOS and 4,000 NLOS samples are collected, for subsequent training.
We also include an open source dataset to prevent the overfitting problem associated with data shortage~\cite{bregar2016nlos}. 
The open dataset contains 42,000 samples\footnote{This LOS/NLOS open datasets were created by using Qorvo DWM1001 UWB radio module. Measurements were taken on 7 different indoor locations. In every indoor location 3,000 LOS and 3,000 NLOS samples were collected.}.
Therefore, a total of 50,000 samples are used for training and testing the CNN-based model.

On the other hand, IMU sensor data is simultaneously collected once every 60~ms.
The IMU sensor data is composed of the gravity and a pure acceleration of x,y, and z coordinates, which means that 6 sensor values are collected at each timestep.
The reason that we collect those two type of data is that the gravity values contain the information about the orientation of the smartphone and the accelerometer values reflect the signature of the user movement, especially during walking~\cite{shu2015last}.
% , zhou2014use}.
Since IMU data resolution is approximately 3 times higher than DS-TWR, we collect 6,600 IMU samples for each pose.
In case of IMU, no open source dataset is used, and hence, in total 26,400 samples are used.

%% file: 4_proposed.tex
\begin{figure}[t]
\centering
\includegraphics[width=1\columnwidth]{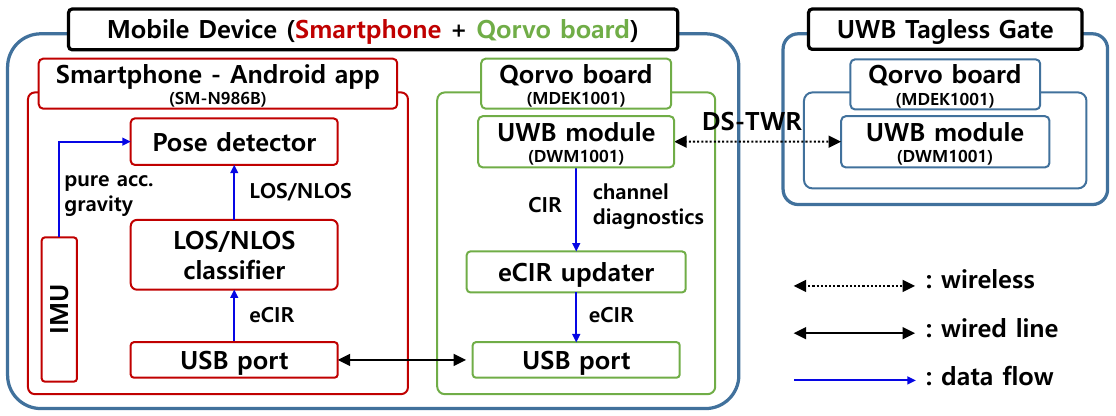}
\vspace{-5mm}
\caption{The overall architecture of {\pp}. MD is a set of the commercial smartphone and the Qorvo board using wired connection.}
\label{fig:architecture}
\vspace{-5mm}
\end{figure}
\section{{\pp}: Proposed system}
\label{sec:proposed}

\subsection{Overview}
\label{subsec:overview}
Fig.~\ref{fig:architecture} shows the overall architecture of {\pp}.
MD is the combination of the smartphone and Qorvo board (i.e., MDEK1001) through USB serial\footnote{The latest smartphones from Samsung, such as Galaxy S22+ and Galaxy Z Fold3, have embedded UWB module, but there is no developer API to obtain CIR yet. 
This is the reason why we choose 3rd party UWB module.}.
The UTG is also equipped with the same UWB module of MD to perform UWB ranging.
After receiving RFM, UWB module on the MD collects CIRs and channel diagnostics, and sends them to the eCIR updater.
The eCIR updater generates eCIR from 1016 CIRs based on channel diagnostics such as maxNoise and FP\_INDEX (Section~\ref{subsec:eCIR}).
After eCIR is generated, it is transferred from Qorvo board to the smartphone through USB port.
The smartphone runs the LOS/NLOS classifier to determine whether the received signal is LOS or NLOS using eCIR whenever eCIR reception occurs (Section~\ref{subsec:los_nlos_classifier}).
The LOS/NLOS classifier sends the result of the classification to the pose detector.
Finally, the pose detector determines the pose %of MD 
by combining IMU sensor data, especially accelerometer and gravity measurements (Section~\ref{subsec:pose_detector}).
In both MD and UTG, each process of all the components operates in real-time.

\subsection{Effective channel impulse response}
\label{subsec:eCIR}

Since {\pp} performs LOS/NLOS classification and the pose detection of MD on Android application, the data should be transferred from the Qorvo board to the smartphone through USB port. 
Therefore, the latency of data transfer between the Qorvo board and the smartphone should be optimized to achieve the real-time operation.
If the processing time of the whole procedures from Qorvo board to smartphone is larger than the ranging interval, the ranging operation between Qorvo board of MD and UTG will be interrupted.

\begin{figure}[t]
    \centering
    \subfigure[The CIR divided into signal and noise based on maxNoise]{
    \includegraphics[width=0.444\columnwidth]{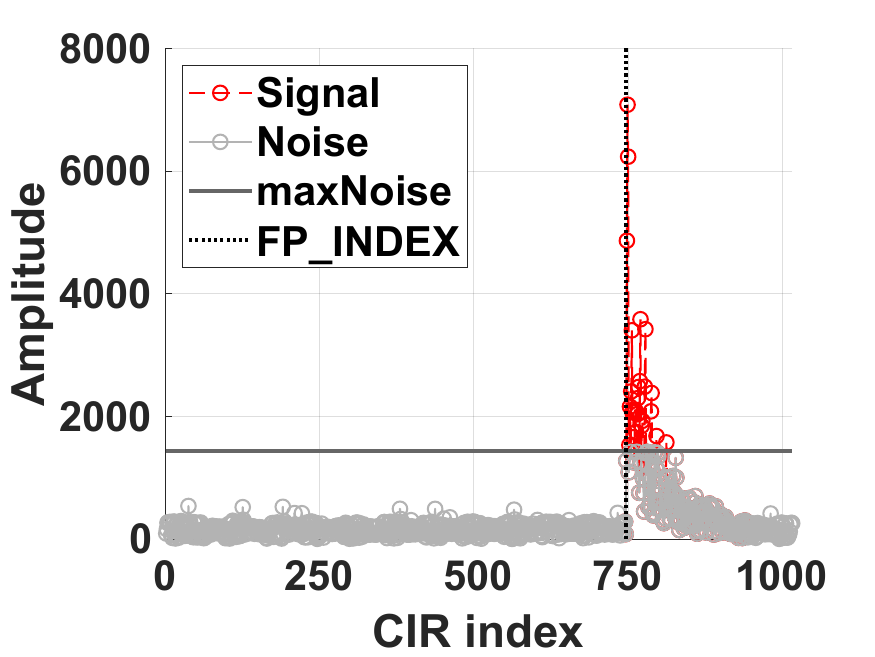}
    \label{fig:cir1016}
    }
    \subfigure[The difference in the amplitude and the number of peaks]{
    \includegraphics[width=0.444\columnwidth]{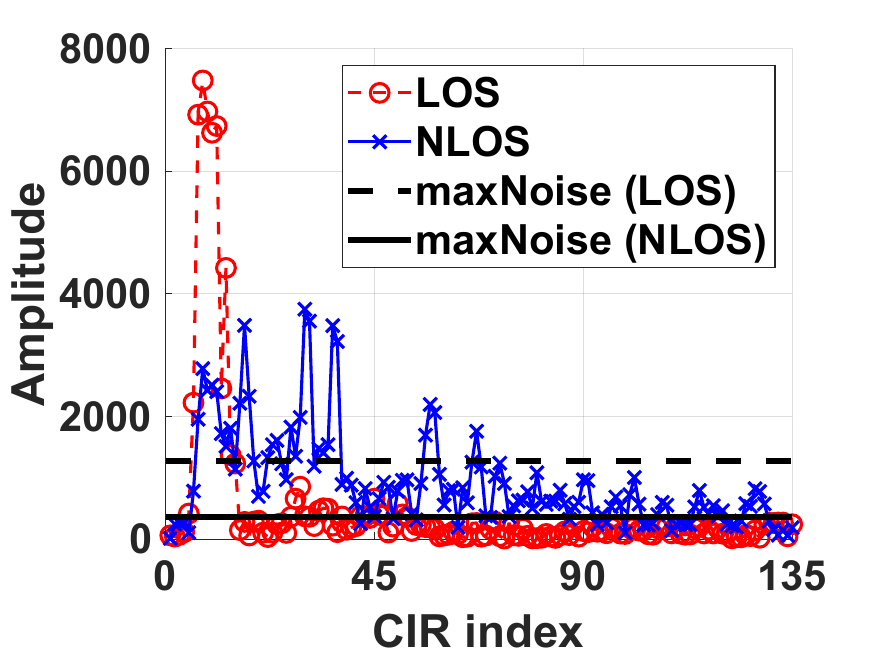}
    \label{fig:cir200}
    }
    \vspace{-2mm}
    \caption{The example of effective CIR and difference between LOS/NLOS.}
    \label{fig:eCIR}
    \vspace{-5mm}
\end{figure}

First, we experimentally measured the latency when 8 channel diagnostics and 1016 CIRs are transferred to the smartphone through USB without data loss, and 223.4~ms is consumed on average, which is longer than the default ranging interval.
Therefore, the latency should be minimized by selective data transfer, and hence, we adopt the eCIR updater to collect the only effective CIRs.
Fig.~\ref{fig:cir1016} represents the CIRs divided into a real signal and a noise using maxNoise.
If the amplitude of CIR is less than maxNoise, the CIR is considered as a noise.
Thus, CIRs before FP\_INDEX are noise and can be eliminated.
Therefore, we focus on CIRs after FP\_INDEX, which contains the real information of the signal.

Fig.~\ref{fig:cir200} shows only 135 CIRs of two LOS and NLOS CIR samples.
The 0 index is FP\_INDEX-5 index as a margin for FP\_INDEX in both CIR samples. 
The CIR trends have clear differences between LOS and NLOS in the max amplitude of the peak and the presence of multiple peaks after the first peak.
Those signatures represent the nature of LOS and NLOS signals.
In the case of LOS, the amplitude of CIRs quickly goes below under maxNoise, and 61 CIRs are actual signal after FP\_INDEX on average.
However, the first peak amplitude of the NLOS signal is much smaller than that of LOS and multiple peaks are observed after the first peak because of the multipath signals in NLOS situation.
Therefore, 130 CIRs are larger than maxNoise after FP\_INDEX in case of NLOS on average.
As a result, at least 135 CIRs (i.e., 130 and 5 margin CIRs before FP\_INDEX) should be transferred to the smartphone for LOS/NLOS classification since those CIRs contain all the signatures of the received signal.
We consider those 135 CIRs as eCIR and using only 135 CIRs for LOS/NLOS classification instead of 1016 CIRs, and hence, the latency is reduced to 17.8~ms from 223.4~ms, which is much smaller than the ranging interval.
 
\subsection{LOS/NLOS classifier}
\label{subsec:los_nlos_classifier}

\begin{figure}[t]
    \centering
    \subfigure[LOS/NLOS classification model architecture with 4 CNN layers]{
    \includegraphics[width=0.9\columnwidth]{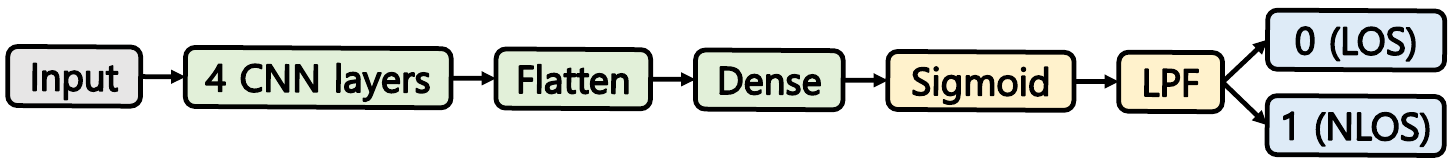}
    \label{fig:nlos_classifier_architecture}
    }
    \subfigure[CNN layer architecture]{
    \includegraphics[width=0.9\columnwidth]{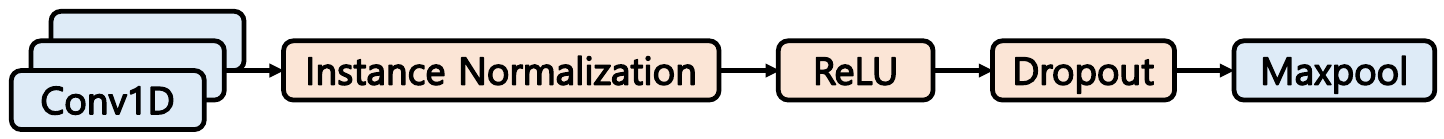}
    \label{fig:nlos_classifier_cnn}
    }
    \vspace{-2mm}
    \caption{CNN based model for LOS/NLOS classification using eCIR input.}
    \label{fig:nlos_classifier}
\end{figure}

Fig.~\ref{fig:nlos_classifier_architecture} shows the deep learning model architecture adopted for the CIR based LOS/NLOS classification. 
The input layer comprises the 135 consecutive CIRs from FP\_INDEX-5. 
It is followed by 4 CNN layers, which are highly efficient in extracting meaningful practical features from spatial/temporal data. 
The output of the CNN is then flattened from the 2D image format to a 1D vector, using the Flatten layer. 
This is followed by a Dense layer, in which all the neurons of the previous and the current hidden layer are pairwise connected using weights. 
The output of the Dense layer is passed into the Sigmoid activation layer, which applies Sigmoid function to produce a number between 0 and 1, which in this case indicates the probability of the CIR corresponding to a NLOS condition. 
Finally, the output of the Sigmoid is smoothed by the low-pass filter~(LPF) to remove the outliers, and the signal is classified as 0 (LOS) or 1 (NLOS). 

The detailed architecture of an individual CNN layer is shown in Fig.~\ref{fig:nlos_classifier_cnn}. 
The convolution layer Conv1D consists of a pre-specified number of 1D vectors called filters, all of them having a pre-specified size called kernel size. 
These filters help to extract significant spatial/temporal features from the input data. Each of these filters is convolved with the input data to produce a vector. 
These output vectors are stacked one above another to give a 2D image type output. 
The output of the convolution layer is passed into the Instance Normalization (IN) layer, which applies normalization per data sample along a specified axis. 
Unlike the conventional Batch Normalization layer, IN thus reduces incoming noise and model dependence on the training data statistics. 
The normalized output is sent into the rectified linear units~(ReLU) layer, which applies the ReLU activation function.
This is followed by the Dropout layer, which randomly drops a fraction of the neurons at algorithm runtime. 
This reduces the overfitting error of the model to the training data. 
The final layer is a Maxpool layer, which also operates based on a pre-specified kernel size. 
It does the maxpooling operation (i.e. chooses the maximum element along a specified axis) in the pre-specified kernel size, and replaces the elements in that kernel with this chosen maximum element. 
This helps to bring down the dimensionality of the data and helps to simplify the model and prevent overfitting. 
The entire set of parameters used in the CNN based model for the LOS/NLOS classification are listed in Table~\ref{tab:NLOS}.
\begin{table}[t]
\vspace{-3mm}
\centering
    \caption{The set of LOS/NLOS classification model parameters}
    \vspace{-2mm}
    \begin{tabular}{cc}
    \toprule[0.6pt]\midrule[0.3pt]
    Model parameters & Value \\
    \midrule
    % No. of CNN layers & 4 \\
    Conv1D layer [1,2,3,4] kernel size & [5,11,17,5]  \\
    Conv1D layer [1,2,3,4] no. of filters & [64,128,256,512] \\
    Dropout rate & 0.2 \\
    Maxpool layer kernel size & 2 \\
    Optimizer & Adam \\
    Loss function & Binary cross-entropy \\
    Batch size & 50 \\
    Maximum no. of epochs & 20 \\
    \midrule[0.3pt]\bottomrule[0.6pt]
  \end{tabular}
  \label{tab:NLOS}
  \vspace{-5mm}
\end{table}

\subsection{Pose detector}
\label{subsec:pose_detector}
% IMU difference between different poses

\begin{figure}[t]
    \centering
    \subfigure[LOS condition]{
    \includegraphics[width=0.444\columnwidth]{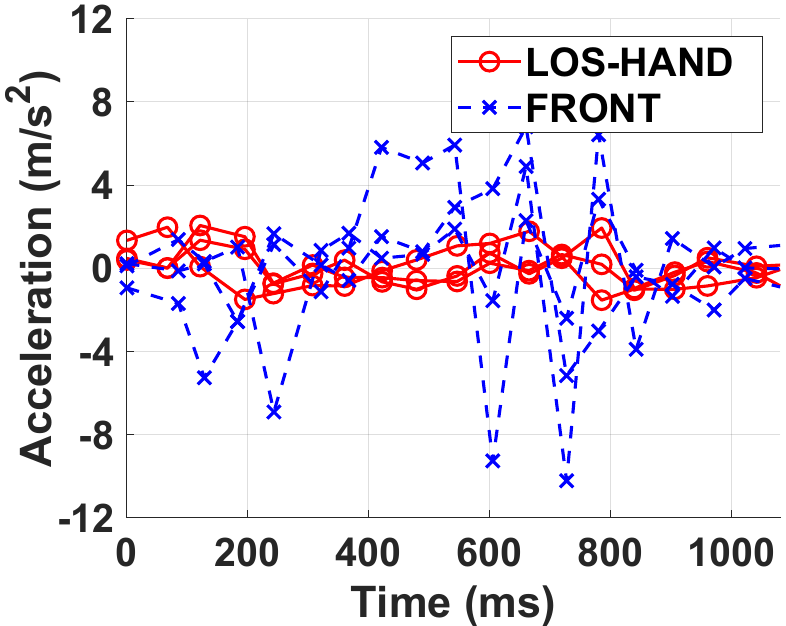}
    \label{fig:imu_los}
    }
    \subfigure[NLOS condition]{
    \includegraphics[width=0.444\columnwidth]{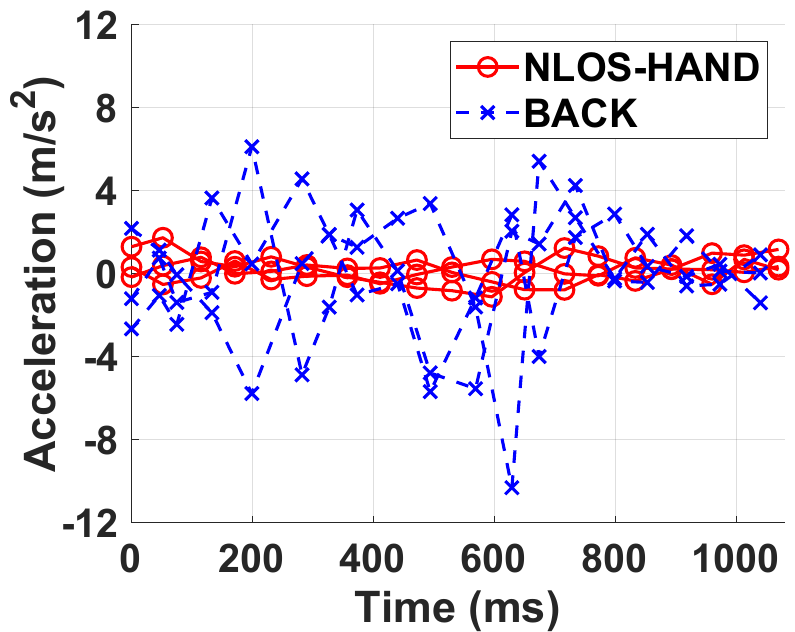}
    \label{fig:imu_nlos}
    }
    \vspace{-2mm}
    \caption{The snapshots of 18 accelerometer values (1080~ms) of four poses.}
    \label{fig:imu_diff}
    \vspace{-5mm}
\end{figure}

Fig.~\ref{fig:imu_diff} shows the snapshots of IMU sensor values, especially accelerometer of each poses, divided into LOS and NLOS conditions. 
The red and blue plot represent the poses of MD, respectively, and three lines of each pose represents the acceleration value of x,y, and z coordinates.
The trends of acceleration value show clear difference whether pose is HAND or POCKET.
If the pose of MD is close to the ground such as POCKET, the magnitude of accelerometer is much higher than that of HAND.
Therefore, the difference in acceleration is a signature of HAND and POCKET, and can be classified by deep neural network.

\begin{figure}[t]
    \centering
    \subfigure[Pose detection model architecture with 3 CNN layers]{
    \includegraphics[width=0.95\columnwidth]{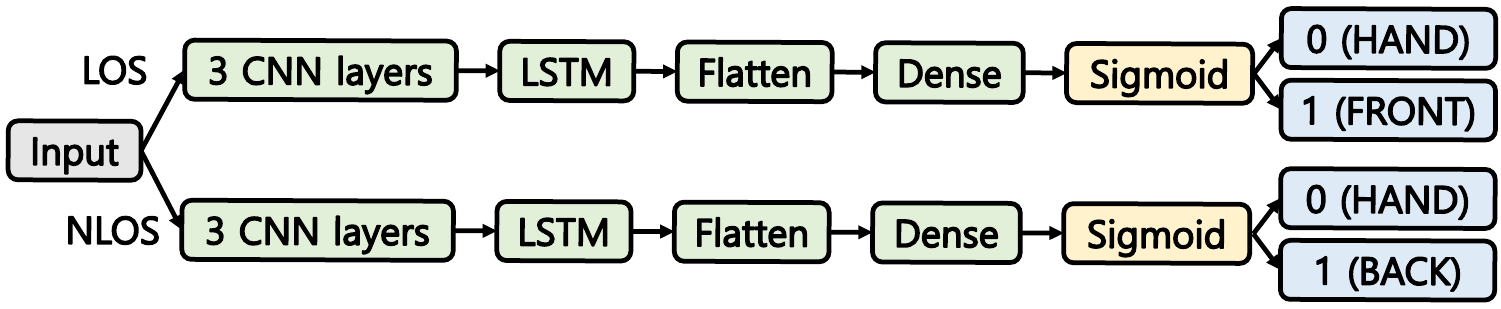}
    \label{fig:pose_classifier_architecture}
    }
    \subfigure[CNN layer architecture]{
    \includegraphics[width=0.95\columnwidth]{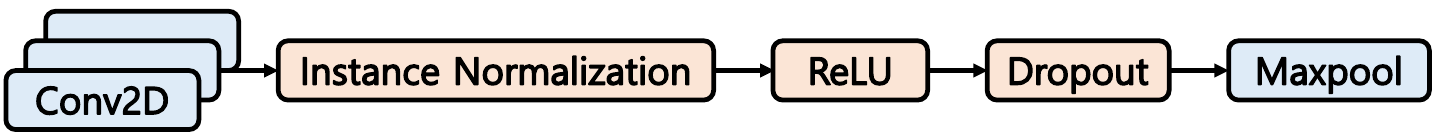}
    \label{fig:pose_classifier_cnn}
    }
    \vspace{-2mm}
    \caption{CNN-LSTM based model for pose detection using IMU input based on LOS/NLOS classification model output.}
    \label{fig:pose_classifier}
    % \vspace{-2mm}
\end{figure}

Fig.~\ref{fig:pose_classifier_architecture} shows the deep neural network model used for pose classification. 
The input layer comprises the IMU sensor data with 6 features, namely the accelerometer and gravity sensor readings along x,y and z coordinates. 
We also accumulate this IMU sensor data for the previous 18 timesteps, which leads to the input being a matrix of dimensions $18\times6$. 
This number of timesteps is derived based on the logic that we want to send the sequential data of the previous 2-3 steps of the user as input data. 
Considering that the average human walking speed is 3 steps per second, that corresponds to sending 1~s worth of IMU sensor data~\cite{shu2015last}. 
The IMU sensor collects data every 60~ms, hence, to collect the IMU data for 1~s, we need to collect 18 IMU data samples. 

The IMU input data is passed into one of two CNN-LSTM models based on whether the CIR based LOS/NLOS classification in the previous stage produced an output LOS or NLOS. 
The two CNN-LSTM based models are designed identically but get trained with different sets of training data, resulting in different optimal weights and filters. 
The model starts with 3 CNN layers, followed by the LSTM layer. 
While the CNN layers extract meaningful spatial/temporal features from the IMU data, the LSTM layer takes advantage of the sequential (18 timesteps) data to produce predictions based on current timestep as well as the previous timesteps~\cite{hochreiter1997long}.
The LSTM output is flattened using a Flatten layer, and then input to a Dense Layer. 
Finally, Sigmoid activation layer operates on the previous layer output, to produce a number between 0 and 1. 
The output of Sigmoid is directly used for pose detection without passing through LPF because LOS/NLOS classification results are already filtered by LPF.

For the model corresponding to the LOS case, the output of Sigmoid corresponds to the probability of MD being FRONT, as opposed to being LOS-HAND.
Similarly, for the NLOS case, the output of Sigmoid corresponds to the probability of MD being BACK, as opposed to being NLOS-HAND.
The detailed CNN layer architecture used for the pose detection models is shown in Fig.~\ref{fig:pose_classifier_cnn}. 
The convolution layer Conv2D contains 2D filters of pre-specified kernel size. 
The output of the Conv2D layer is successively passed through an IN layer, ReLU activation, Dropout, and a Maxpool layer. 
The entire set of parameters used in CNN-LSTM based model for the pose classification are listed in Table~\ref{tab:pose}.

\begin{table}[t]
\vspace{-2mm}
\centering
    \caption{The set of pose detection model parameters}
    \vspace{-2mm}
    \begin{tabular}{cc}
    \toprule[0.6pt]\midrule[0.3pt]
    Model parameters & Value \\
    \midrule
    % No. of previous timesteps & 18 \\
    % No. of CNN layers & 3  \\
    Conv2D layer [1,2,3] kernel size & [2,2,2]  \\
    Conv2D layer [1,2,3] no. of filters & [64,128,256] \\
    Dropout rate & 0.2 \\
    Maxpool layer kernel size & 2 \\
    LSTM layer no. of units & 128 \\
    Optimizer & Adam \\
    Loss function & Binary cross-entropy \\
    Batch size & 100 \\
    Maximum no. of epochs & 100 \\
    \midrule[0.3pt]\bottomrule[0.6pt]
  \end{tabular}
  \label{tab:pose}
  \vspace{-5mm}
\end{table}

%% file: 5_evaluation.tex
\section{Performance Evaluation}
\label{sec:evaluation}

\subsection{Implementation}
\label{subsec:implementation}
\noindent\textbf{MD:}
We implemented the application side of {\pp} on the commercial smartphone (i.e., SM-N986B) running on Android~11 to show the feasibility.
Both LOS/NLOS classifier and pose detector are implemented using TensorFlow 2.10.0 and converted to TensorFlow-Lite for mobile deployment.
Other modules including CNN models and IMU are implemented using Android Studio.
The exponentially weighted moving average is used as a LPF with a weighting factor of 0.8.
% ~\cite{hunter1986exponentially}.
If the user turns on {\pp}, the results of LOS/NLOS classification and pose detector are showed on application UI.
% in real-time.

\noindent\textbf{UWB:}
We implemented both UWB modules of MD and UTG based on the example code from the Qorvo DWM1001~\cite{dwm1001example}. 
The DS-TWR operation between MD/UTG and eCIR updater in Qorvo board is implemented by following the user manual using SEGGER Embedded Studio for ARM 5.32a~\cite{dw1000user, seggerARM}.
% All data generated from the UWB module and IMU sensor are smoothed out by using low-pass filter to remove the high frequency noise caused by the random movement of the user.
The UWB module is connected to MD via a wired connection.

\subsection{Experiment setup}
\label{subsec:experiment}
\noindent\textbf{Datasets and CNN training:} We trained our LOS/NLOS classifier on open datasets and collected data with preliminary experiments to prevent overfitting, and pose detector is trained by using only collected IMU data.
All datasets were randomly divided into training (80\%) and testing (20\%) data.

\begin{figure}[t]
\centering
\includegraphics[width=0.92\columnwidth]{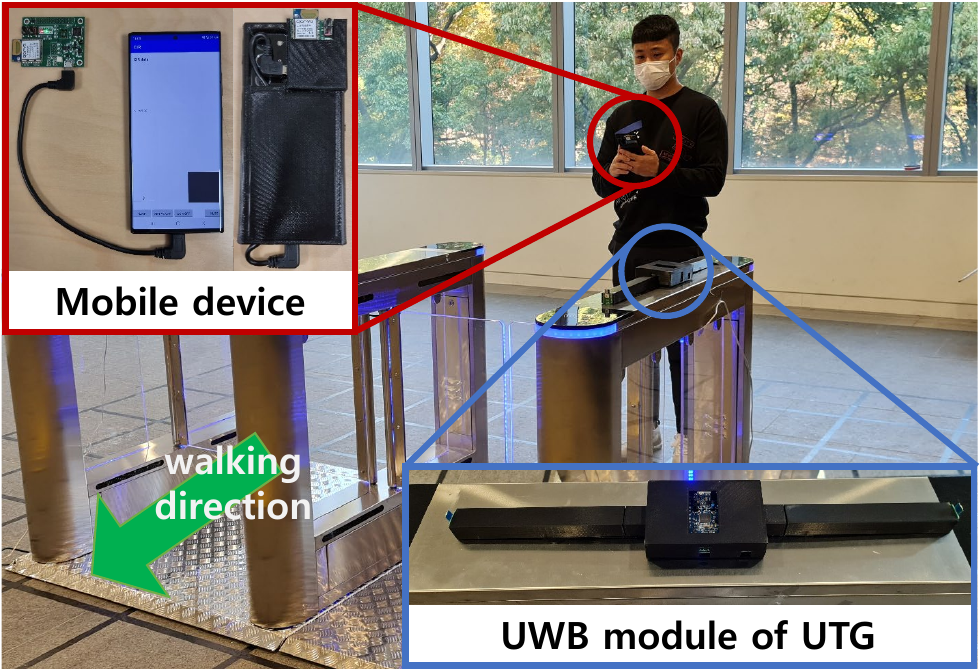}
\vspace{-2mm}
\caption{The snapshot of the real-world UTG experiment environment.}
\label{fig:real-world}
\end{figure}

\noindent\textbf{Performance evaluation:}
We conducted our experiments for a test area of $8\times8~m^2$ in an office building.
As shown in Fig.~\ref{fig:real-world}, UTG is installed with UWB module and MD is composed of the smartphone and UWB module.
We also design the case of MD using 3D printer for the consistency of experiments.
The experiments are focused to show the real-time accuracy of LOS/NLOS classification and pose detection, and hence, we iterate 50 times for each poses.
Each iteration returns the LOS/NLOS condition and pose of MD every 200~ms while the user with MD starts walking towards UTG from 4~m away.

\subsection{Performance}
\label{subsec:performance}
\begin{table}[t]
\vspace{-2mm}
\centering
    \caption{The accuracy of LOS/NLOS classification and pose detection.}
    % \caption{The real-time performance of LOS/NLOS classification and pose detection accuracy.}
    \vspace{-2mm}
    \begin{tabular}{c|c|c}
    \toprule[0.6pt]\midrule[0.3pt]
     & {LOS/NLOS classification (w/o LPF)} & {Pose detection} \\
    \midrule
    {LOS-HAND} & 0.994~~(0.849) & 0.983 \\
        \midrule
    {NLOS-HAND} & 0.993~~(0.839) & 0.982 \\
        \midrule
    {FRONT} & 0.968~~(0.783) & 0.931 \\
        \midrule
    {BACK} & 0.982~~(0.803) & 0.948 \\
    \midrule[0.3pt]\bottomrule[0.6pt]
  \end{tabular}
\label{tab:performance}
\vspace{-5mm}
\end{table}

We measure the LOS/NLOS classification and pose detection performance of {\pp} as the rate of true positive an false positive, in which the classification and detection are correctly/wrongly conducted.
Table~\ref{tab:performance} summarizes the accuracy of each pose measured in real-time.

\noindent\textbf{LOS/NLOS classification:}
The two values are obtained from experiments.
The first value is the actual performance of LOS/NLOS classification model and the second value in the parentheses represents the accuracy without LPF, which is the output of Sigmoid in Fig.~\ref{fig:nlos_classifier_architecture}.
The performance without LPF is similar to that of \cite{jiang2020los}, which proposes CNN-LSTM model for classification, and overall accuracy of 0.819.
{\pp} improves the accuracy by adding LPF to CNN model, and the overall LOS/NLOS classification accuracy is 0.984.
As a result, {\pp} improved classification performance by 20.1\%. 

\noindent\textbf{Pose detection:}
Considering that the accuracy of pose detection is affected by the result of LOS/NLOS classification, the pose detection accuracy can not outperform LOS/NLOS classification accuracy.
The average performance of pose detection is 0.961, and there is only accuracy degradation of 0.023 compared to LOS/NLOS classification.
Therefore, we conclude that {\pp} successfully classifies four different poses.
% based on the eCIR and IMU.

\noindent\textbf{Pose transition delay:}
{\pp} utilizes LPF to improve the accuracy of LOS/NLOS classification,
% , and gains accuracy improvement.
but, LPF has trade-off between the accuracy and a delay.
The higher the weighting factor, the longer delay.
Therefore, we evaluate the pose transition delay between LOS and NLOS condition.
To this end, we record the timing of the transition and measure the delay using application.
If {\pp} does not adopt LPF on LOS/NLOS classification, the average pose transition delay should be 200~ms, which is same as the default ranging interval of DS-TWR.
Table~\ref{tab:delay} summarizes the average transition delay and standard deviation for the possible combinations in ms.
The transition delay between HANDs takes 734.8~ms, which is the shortest among four delays, because it has no change of the position on the human body.
The remaining transitions contain pose change from/to POCKET that have more trajectory to move on the body, and consume additional 111.1~ms on average.
All pose transitions are completed within one second.

\begin{table}[t]
% \vspace{-2mm}
\centering
    \caption{The transition delay between LOS and NLOS conditions.}
    \vspace{-2mm}
    \begin{tabular}{c|c|c}
    \toprule[0.6pt]\midrule[0.3pt]
     & LOS-HAND & FRONT\\
    \midrule
    {NLOS-HAND} & 734.8~~(137.4) & 838.4~~(168.9) \\
    \midrule
    {BACK} & 848.0~~(173.8) & 851.2~~(188.4) \\
    \midrule[0.3pt]\bottomrule[0.6pt]
  \end{tabular}
\label{tab:delay}
\vspace{-4mm}
\end{table}

%% file: 6_conclusion.tex
\section{Conclusion and future work}
\label{sec:conclusion}
In this paper, we have proposed {\pp}, a novel pose detection system for the proximity services based on UWB CIR and fusion of IMU sensors.
We implemented and evaluated {\pp} in a real-world environment using SM-N986B and MDEK1001.
{\pp} achieved {\cc} and {\dt} accuracy in LOS/NLOS classification and pose detection in real-time.
As a future work, we will design the algorithm for alleviating ranging error in UWB by adopting {\pp}.
In addition, we have a plan to develop the whole UTG system, and will implement {\pp} as a one device without external UWB module. 

% DL-TDoA Extension (localization accuracy를 향상시킬 수 있을 것)
% Pose detection을 실제로 적용해서 Tagless Gate의 performance를 향상시키는 것
% 폰에서 API를 제공해서 CIR을 얻을 수 있다면, external board를 활용하지 않고 구현할 수 있음